\documentstyle[openbib,12pt,fleqn]{article}
\newcommand{\bea}{\begin{eqnarray}}
\newcommand{\beq}{\begin{equation}}
\newcommand{\eea}{\end{eqnarray}}
\newcommand{\eeq}{\end{equation}}

\renewcommand{\baselinestretch}{2}

\begin{document}
\bibliographystyle{unsrt}

\setcounter{footnote}{0}

%
%
%
%
\begin{center}
\phantom{.}
{\Large \bf Few interacting particles in a random potential \\}
{\small \sl   D.L.SHEPELYANSKY~$^{(a)}$ and
 O.P.SUSHKOV~$^{(a,b)}$ \\}
{\small \it Laboratoire de Physique Quantique,
  Universit\'e Paul Sabatier, \\
  118, route de Narbonne,31062 Toulouse Cedex, France}\\

\vspace{0.5truecm}

\vskip .3 truecm

\vspace{0.5truecm}
\end{center}
\small
{\bf Abstract:\/}
We study the localization length of few interacting particles in a 
random potential. Concentrating on the case of three particles we
show that their localization length is strongly enhanced 
comparing to the enhancement for two interacting particles.

\vskip .6 truecm


\newpage


Recently it had been shown that in a random potential
two repulsing/attracting particles can propagate coherently
on a distance $l_c$ which is much larger than one particle
localization length $l_1$ in absence of interaction \cite{TIP}.
In some sense interaction destroys quantum interference 
which leads to one-particle localization and creates
an effective pair of two particles of size $l_1$ propagating on a 
large distance. For better understanding of this result
Imry developped \cite{Imry} a scaling block picture of localization for
interacting particles which can be applied in principle for
a larger number of particles and higher dimensions.
Intensive numerical investigations of Pichard and coworkers
\cite{Pichard} and von Oppen and coworkers \cite{Oppen}
confirmed the existence of the two interacting particles (TIP) effect.
While some additional checks are still required the results
\cite{Pichard,Oppen} definitely show that in one-dimensional case
the TIP length is $l_c \propto {l_1}^\alpha$ with $\alpha$ close
to the theoretically predicted power \cite{TIP} $\alpha=2$.
These results are also in agreement with the previous studies
of Dorokhov who analysed the case of two particles confined
by strong attraction in a well with size much smaller than $l_1$
\cite{Dorokhov}. 
The investigations of TIP effect in higher dimensions have been
done in \cite{Imry,Borg1,Frahm3d} and they demonstrated that
in dimension $d=3$ the TIP pair can be delocalized below
one-particle Anderson transition where all one-particle states are localized.

While now the properties of TIP propagation reached a level of qualitative
understanding, the problem of a larger number of interacting particles 
is still not well understood. From the physical point of view 
the most interesting situation is the case of finite
density of particles. However, the analysis in this case is quite complicated
and at present only estimate \cite{Imry} and numerical
studies in \cite{Oppen1} have addressed this problem.
One of the ways to approach this problem is to analyse the case
of larger number of particles. The simplest case is 
three interacting particles where the situation is not
so trivial since two particle interaction leads to 
the Breit-Wigner structure of eigenstates \cite{Jaq,Frahm1,Feod,Pich1}.
In this paper we will concentrate on this three particle model.

However, before the analysis of three particle model let us first 
discuss some generalized model of TIP (see also \cite{Moriond}).
In this model the first particle is moving on sites  $n_1$ in a one channel
Anderson model with diagonal disorder changing in the interval
$\pm W_1$, the intersites hopping matrix element is $V_1$ and localization
length at the center of the band is $l_1 \approx 25 (V_1/W_1)^2 \gg 1$.
The second particle is moving in a strip with $M$ transverse channels
with sites marked by index $n_2$ along the strip and index ${\tilde{n}_2}
(1 \leq {{\tilde{n}}_2} \leq M) $ in transverse direction. The disorder
in the strip is independent of disorder in the chain with first particle
and the localization length for the second particle  is $l_2 \propto M$.
The hopping in the strip is $V \approx V_1$, and we assume that $l_2 > l_1$.
Now we will analyse what will happen if the interaction of the
form $U \delta_{n_1,n_2}$ is switched on between two particles.
Similarly to \cite{TIP} one should first estimate
the transition matrix elements $U_s$ between eigenstates 
without interaction ( $U=0$).
This gives
\beq
{U_s} = U {\sum_{{n_1},{n_2}, {{\tilde {n}}_2}} }
R^{+}_{n_1, m_{1}} 
{\tilde{R}}^{+}_{n_2,{\tilde {n}}_2, m_{2},{{\tilde {m}}_2}} 
R_{n_1, m^{'}_{1}} 
{\tilde{R}}_{n_2,{\tilde {n}}_2, m^{'}_{2},{{\tilde {m'}}_2}} 
\delta_{n_1,n_2} 
\label{us}
\eeq
where $R$ represents the transformation between the lattice basis
and one-particle eigenstates so that $R_{n_1,m_1} \approx
\exp(-{\mid {n_1 - m_1} \mid}/{l_{1}}-{i}\theta_{n_1,m_1})/{\sqrt{l_{1}}}$
and $\tilde{R}_{n_2,{\tilde{n}}_2,m_2,{\tilde{m}}_2} \approx $\\
$\exp(-{\mid {n_2 - m_2} \mid}
/{l_{2}}-{i}\theta_{n_2,{\tilde{n}}_2,m_2,{\tilde{m}}_2})
/{\sqrt{M l_{2}}}$ correspondingly for the first and second particle.
The phase $\theta$ randomly changes with indices.
Due to the exponential decrease of $R$
one should take into account only the states with $\mid n_{1,2} - m_{1,2} \mid
< l_{1,2}$. For the case $l_2 > l_1$ the sum in (\ref{us}) 
containes approximately $l_1 M$ random terms so that 
$U_s \approx U/(l_2 \sqrt{l_1 M})$. The interaction induced transition rate
is given by the Fermi golden rule $\Gamma \sim {U_s}^2 \rho_c $
where $\rho_c \approx l_1 l_2 M/V$ is the density of coupled states.
As the result, $\Gamma \sim U^2/(V l_2)$ is independent of $l_1$ and $M$.

With the rate $\Gamma$ we can determine the interaction induced diffusion
rate for the first particle which is $D_1 \sim {l_1}^2 \Gamma 
\sim U^2 {l_1}^2/(V l_2)$ and appears as the result of collisions of
the first particle with the second one oscillating in the 
block of size $l_2$.
Knowing the diffusion rate, it is possible to
determine the localization length for a pair in a way similar
to that used for the kicked rotator \cite{PhysD87} and 
based on the uncertainty relation between the frequency and time
(see also \cite{Moriond}). Indeed, the number of excited states
in the first chain grows with time $t$ as $\Delta n_1 \sim (D_1 t)^{1/2}$.
Since two particles are propagating together so that 
$\mid n_1 - n_2 \mid < l_2$ the total number of excited states in both
chains is $\Delta N \sim \Delta n_1 (M l_2) \delta E/V$
where $\delta E$ takes into account the factor that the 
states are excited only in some energy interval inside the band width $V$.
Generally, $\delta E < V$ and it is of the order of Breit-Wigner width $\Gamma$
\cite{Jaq},
but we will see that $\delta E$ does not enter in the final expression for 
the localization length of pair, and therefore actual value of $\delta E$
is not very important (see also \cite{PhysD87}). Indeed, all these
$\Delta N$ levels are homogeneously distributed in the 
energy interval $\delta E$ and the average splitting between 
them is $\Delta \nu \sim \delta E/\Delta N$. 
According to the uncertainty relation between frequency and time
at the moment $t$ we can resolve discrete lines with the
splitting $1/t$. 
Therefore, at the moment $t^*$ 
defined by the equation $\Delta \nu \sim 1/t^*$
the discreteness of the spectrum is resolved and 
the diffusive propagation is stopped at
$t^* \sim \Delta N(t^*)/\delta E$. This condition gives
the localization time $t^*$ for TIP pair and the localization length
for the first particle $l_{c1}$:
\beq
t^* \sim U^2 {l_1}^2 M^2 l_2/V; \;\;\; 
l_{c1} \sim \Delta n_1 \sim (U/V)^2 {l_1}^2 M
\label{loc}
\eeq

The interesting feature of this result is that $l_{c1}$ is independent
of $l_2$. This indicates that the nature of motion in the 
second chain does not influence much the localization in the first chain.
For $M=1$ the length $l_{c1}$ is the same as in the case
of TIP localization in one chain. However, the growth
of the number of channels $M$ in the strip leads to the increase
of $l_{c1}$. The localization length for the second particle
is $l_{c2} \sim l_2$ if $l_2 \gg l_{c1}$ and $l_{c2} \sim l_{c1}$
if $l_2 \ll l_{c1}$. The similar  approach can be used 
for analysis of TIP localization in higher dimensions \cite{Moriond}.

Let us now consider three interacting particles in 
one-dimensional chain with on site interaction
$U_{12} \delta_{n_1,n_2}$, $U_{23} \delta_{n_2,n_3}$
and $U_{13} \delta_{n_1,n_3}$ where $n_{1,2,3}$ marks the
site position of corresponding particle in the chain.
As above, the one-particle localization length is $l_1$ 
and the band width is $4V$. For simplicity we will assume that
$U_{13}=0$ and $U_{23} > U_{12}$.
Then in first approximation the particles 2-3 form a pair of size $l_1$
which is localized on the length $l_{c2} \sim (U_{23}/V)^2 {l_1}^2$.
When this pair approaches the first particle
at a distance $l_1$ the interaction between three
particles in a block of size $l_1$ gives mixing between ${l_1}^3$
3-particle states. 
An effective matrix element $U_{s1}$ of interaction 
between 3-particle states in the block of size $l_1$ should been 
calculated in  the second order perturbation theory, since direct interaction
couples only 2-particle states. Therefore, the matrix element
between initial state $|123>$ and final state $|1'2'3'>$
is given by diagram presented in Fig.1 with intermediate state $|1'{\bar 2} 3>$.
It is of the form
\beq
U_{s1} = {\sum_{\bar 2}} 
{{<12| U_{12} |1' {\bar 2}> < {\bar 2} 3| U_{23}| 2' 3'>} \over
{(E_1+E_2+E_3 - E_{1'}-E_{\bar 2}-E_3)}} \sim {{U_{12} U_{23}} \over
{{l_1}^3 \Delta_1}}
\label{pbt}
\eeq
It is important that the summation is carried out only over single particle
states $\bar 2$, hence $\Delta_1 \sim V/l_1$ is 
single particle level spacing. Finally this gives the mixing
rate in a block of size $l_1$ 
\beq
\Gamma_1 \sim {U_{s1}}^2 \rho_3 \sim (U_{12} U_{23}/V^2)^2 V/l_1
\label{g3}
\eeq
where $\rho_3 \sim {l_1}^3/V$ is the density of 3-particle states in the
block. This $\Gamma_1$  gives the mixing
rate during the collision of the first particle with
the pair 2-3 in the block $l_1$. 
The frequency of such collisions is of the order
of $l_1/l_{c2}$ since from ergodicity the ratio of time of the collision
to the time between collisions is proportional to the ratio of 
volumes. Therefore, the average transition rate for 1-particle 
per unit  time is 
${\tilde \Gamma_1} \sim \Gamma_1 l_1/l_{c2} $.
Such transitions give the diffusion rate of the first 
particle $D_1 \sim {\tilde \Gamma_1} {l_1}^2 \sim {U_{12}}^2/V$ 
since the size of transition
is $l_1$. Similarly to the previous case with two chains the
total number of excited states after time $t^*$ is 
$\Delta N \sim (D_1 t^*)^{1/2} (l_{c2} l_1) \delta E/V$ 
where $\delta E$ is an energy width in which the levels are
mixed. The localization time $t^*$, as previously, is determined from the
condition $\Delta N \sim \delta E t^*$ which gives
\beq
t^* \sim D_1 (l_{c2} l_1)^2/V^2; \;\; \; \; \; \;
l_{c1}/l_1 \sim D_1 l_{c2}/V \sim (U_{12} U_{23}/V^2)^2 {l_1}^2
\label{lc1}
\eeq

For $U_{12} \sim U_{23} \sim U$ the localization length for 
the first particle is enhanced  only if there is an enhancement for
two-particle localization length, namely 
$(U/V)^2 l_1 >1$. This result is quite natural since
for $(U/V)^2 l_1 < 1$ two-particle interaction is too weak and it is
not able to mix three-particle levels. Another limiting case
in (\ref{lc1}) corresponds to $U \sim V$. For such interaction
$l_{c1} \sim {l_1}^3$ which is similar to the case of three particles
trapped in a bag of size $l_1$. 
Indeed, one can consider 3-particle bag model like TIP one \cite{TIP}
with effective number of transverse channels
$M_{ef} \sim l_1$, therefore for the 3-particle bag 
$l_{b3} \sim M_{ef} {l_1}^2 \sim {l_1}^3$.
The same estimate for $l_{b3}$ had been also obtained in \cite{PIm}
basing on the approachs developped in \cite{Dorokhov,Imry}.
In some sense the result (\ref{lc1}) shows that similar to the TIP case 
the "size" and "form" of the bag is not important for the effect.
Let us also mention that the case $U_{23} \sim V$ is similar to
previously analysed model of TIP in the chain and the strip (\ref{loc}).
Indeed, here the third particle gives the effective number of
channels $M \sim l_1$ so that (\ref{lc1}) becomes equivalent to
(\ref{loc}).
Generalization of the result (\ref{lc1}) for $k$ particles
gives the enhancement $l_{ck}/l_1 \sim ((U/V)^2 l_1)^{k-1}$.

For the 3-dimensional case $l_1$ in the enhancement factor 
$(U/V)^2 l_1$ should be
replaced by ${l_1}^3$ \cite{Imry,Borg1,Moriond} so the
delocalization takes place if $((U/V)^2 {l_1}^3)^{k-1} > 1$. 
This means that the delocalization border for few particles coincides 
approximately with that for TIP and therefore it is 
not possible to have propagating cluster with $k>2$ repulsive 
particles. In some sense only TIP pairs are well defined.

We thank J.-L.Pichard and Y.Imry for useful and stimulating 
discussions and communication about their results \cite{PIm}
prior to publication. 
One of us (DLS) gratefully acknowledges the hospitality 
and fine working conditions at the Institut Henri Poincar\'e during 
research on the above problem. Another coauther (OPS)
thanks Laboratoire de Physique Quantique,
Universit\'e Paul Sabatier for hospitality and financial suport during the 
work on this problem.

\vfill\eject

\renewcommand{\baselinestretch} {2}

\vfill\eject
{\bf {Figure captions}}
\vskip 20pt
\begin{description}
{
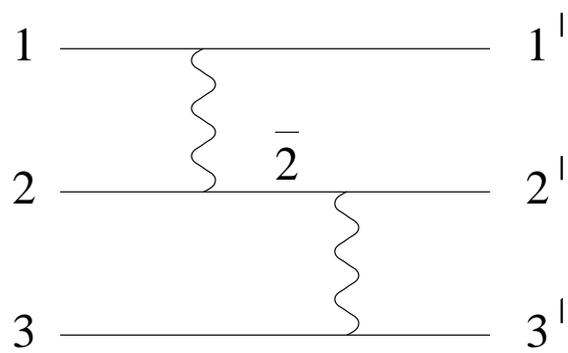
\item[Fig. 1:]  Diagram for effective three particle matrix element
$U_{s1}$ in (3).
}
\end{description}
\end{document}